\documentclass[a4paper,8pt]{article}
\usepackage[cp1251]{inputenc}
\usepackage[russian,english]{babel}
\usepackage{array,longtable}
\usepackage{amssymb, latexsym, amsthm, amsmath, amsxtra}
\usepackage{graphicx}

\graphicspath{{}}
\DeclareGraphicsExtensions{.pdf,.png,.jpg,}

\makeatletter

\begin{document}

sent to Jetp Letters

\large

\begin{center}
\title{}{\bf On Lorentz-invariant 2D equations admitting long-lived localized solutions with a nontrivial structure.  }
\vskip 1cm

\author{}
 R.K. Salimov \textsuperscript{1}, T.R. Salimov \textsuperscript{2}, E.G. Ekomasov \textsuperscript{1}
{}

% author(-s)
\vskip 1cm

{ \textsuperscript{1} Bashkir State University, Ufa, Russia }

{ \textsuperscript{2} Moscow Institute of Physics and Tehnology, Dolgoprudny,  Russia}

%The name of establishment in which research is executed.
%If authors take an identical place of work, the name of the organization is written once.

\vskip 0.5cm
e-mail: salimovrk@bashedu.ru

\end{center}

\vskip 1cm

{\bf Abstract}
 \par
 The article studies Lorentz-invariant 2D equations with long-lived ($t \backsim 1000$ ) localized solutions. In the case of three scalar fields localized solutions with a nontrivial internal structure similar to the hadron structure are showed. In this case, the solutions of the equations are analogous to the confinement model by flux tube.
 \par
 \vskip 0.5cm


\par
\vskip 1cm

\begin{center}
{\bf Introduction}
\end{center}

Soliton solutions of nonlinear equations are often  regarded as extended models of particles[1-9], for example, the Skyrme model [10-12] describing the internal structure of baryons and lightweight nuclei. In the course of this approach particles are described as  a “bunched field” occupying  a bounded region of a space. Such an approach is different from the standard model where the extended particles are constructed of point particles. One of the drawbacks of the soliton approach is  a small number of mathematical models with 2D and 3D stable localized solutions. The work presents Lorentz-invariant 2D equations with long-lived localized solutions.  The solutions have an interesting internal structure the properties of which suggest the analogy with the hadrons’ structure.

\begin{center}
{\bf Results}
\end{center}

 Equations (1,2) with localized  symmetric solutions that do not spread spherically or cylindrically in 3D and 2D  cases have been considered before in [13].

\begin{align}
 u_{rr}+\frac{2u_{r}}{r}-u_{tt}=u^{\frac{1}{2k+1}}\label{eq:1}
   \end{align}

\begin{align}
 u_{rr}+\frac{u_{r}}{r}-u_{tt}=u^{\frac{1}{2k+1}}\label{eq:2}
   \end{align}

Numerical simulation in 2D  case shows that without the condition of cylindrical symmetry the localized solutions of one field are unstable and are quite fast ($t \backsim 30$)  to fall apart. For this reason equations of the form (3-4) for two fields were considered

\begin{align}
 u_{xx}+u_{yy}-u_{tt}=\alpha\frac{u}{(u^2+v^2)^\frac{n}{2n+1}}+\beta u+\gamma u(u^2+v^2)\label{eq:3}
   \end{align}

\begin{align}
 v_{xx}+v_{yy}-v_{tt}=\alpha\frac{v}{(u^2+v^2)^\frac{n}{2n+1}}+\beta v+\gamma v(u^2+v^2)\label{eq:4}
   \end{align}

The equations are interesting because they have long-lived  nonspreading numerical solutions. Localized solutions in them are kept without noticable energy losses during the whole time of the numerical simulation ( $t \backsim 1000$) . A remarkable feature of the solutions with some parameters is their nontrivial structure. Particularly, the solutions represent an oval periodically changing its orientation on plane xy (fig.1 ,2).

\begin{figure}[h]
\center
\includegraphics[width=9cm, height=6cm]{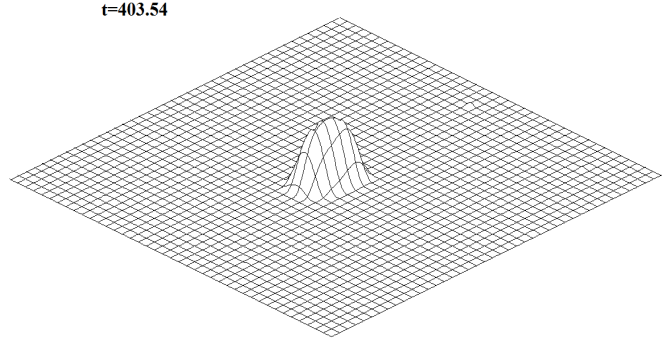}
\caption{Orientation of the solution (3-4) in Ox-direction. $n=8;\alpha=1;\beta=12;\gamma=12$ }
\label{schema}
\end{figure}

\begin{figure}[h]
\center
\includegraphics[width=9cm, height=6cm]{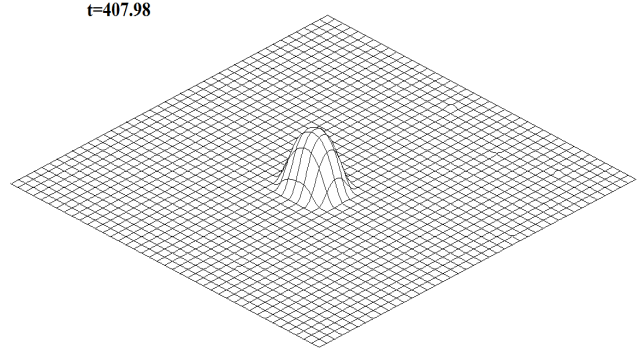}
\caption{Orientation of the solution in Oy-direction.  }
\label{schema}
\end{figure}

   To obtain a more complex  internal structure, solutions of equations (5-7) for three scalar fields were numerically investigated:

\begin{align}
u_{xx}+u_{yy}-u_{tt}=\alpha\frac{u}{(u^2+v^2+w^2)^\frac{n}{2n+1}}+\beta u+\gamma u(u^2+v^2+w^2)+\lambda uv^2
+\xi w
\end{align}

\begin{align}
v_{xx}+v_{yy}-v_{tt}=\alpha\frac{v}{(u^2+v^2+w^2)^\frac{n}{2n+1}}+\beta v+\gamma v(u^2+v^2+w^2)+\lambda vu^2
-\xi w
\end{align}

\begin{align}
w_{xx}+w_{yy}-w_{tt}=\alpha\frac{w}{(u^2+v^2+w^2)^\frac{n}{2n+1}}+\eta w+\gamma u(u^2+v^2+w^2)+\xi (u-v)+
\nonumber\\+\mu \frac{w}{(w^2)^\frac{n}{2n+1}}
\end{align}

                (5-7)

The equations are constructed in such a way that two scalar fields $u ,v$ repel at large amplitudes because of summands   .  The presence of several fields, simultaneously nonzero, results in a more stable localization of solutions. The example of the existence of the solutions of equations (3-4) proves that. Numerical solutions of equations (5-7) have a more remarkable internal structure. Under some initial conditions and parameters ($n=8;\alpha=1;\beta=12;\gamma=18;\lambda=360;\xi=1.2;\eta=8\pi;\mu=1$  ) ,  the solution is a structure of two periodically emerging  maxima of the value of function $(u^2+v^2+w^2)^{(1/2)})$ . See fig. 3.

\begin{figure}[h]
\center
\includegraphics[width=8cm, height=5cm]{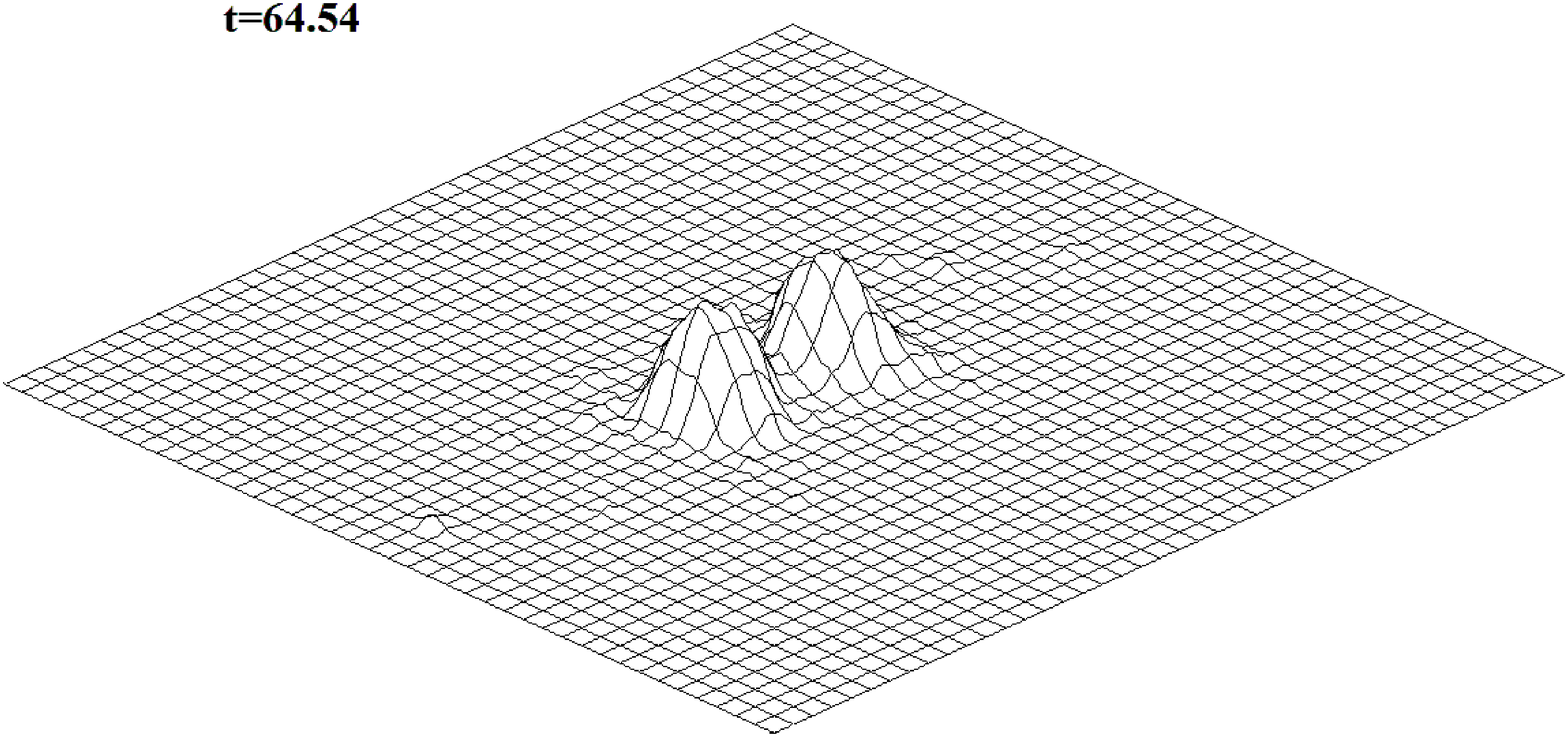}
\caption{  }
\label{schema}
\end{figure}

The solution with a maximal amplitude is changed to the solution with a minimal amplitude of value $(u^2+v^2+w^2)^{(1/2)}$  .  See fig. 4

\begin{figure}[h]
\center
\includegraphics[width=8cm, height=5cm]{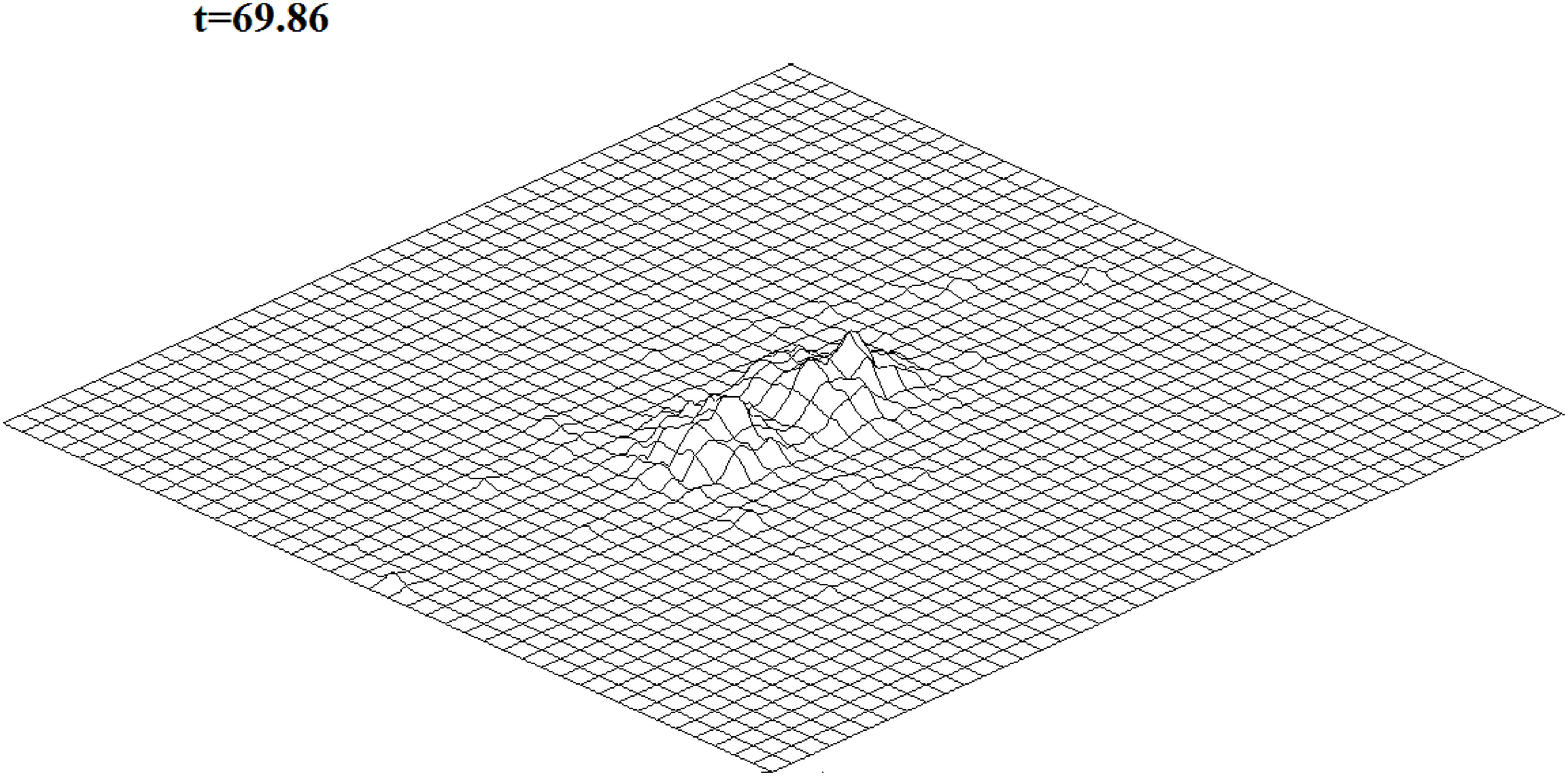}
\caption{  }
\label{schema}
\end{figure}

\begin{figure}[h]
\center
\includegraphics[width=8cm, height=5cm]{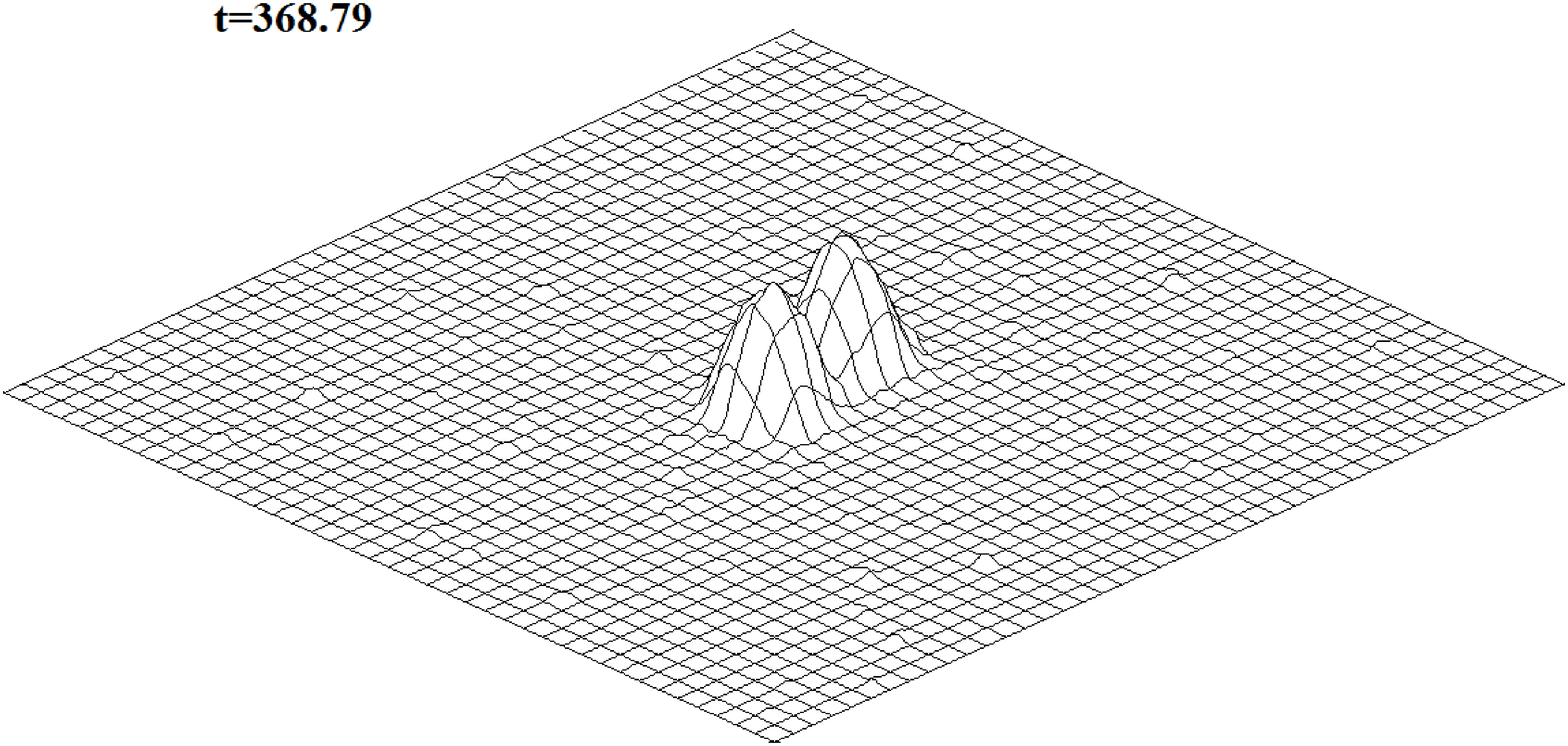}
\caption{  }
\label{schema}
\end{figure}

Such a behavior of the solutions is kept constant for quite a long time ($t \backsim 400$)   without a noticeable distortion of the solution form.

\begin{center}
{\bf Results and discussion}
\end{center}
As is seen from the numerical solution of equations (3-4)  the equations of the scalar fields with fractional nonlinear nature are of interest for being the equations with nonspreading localized solutions.  The existence of deviation of spatial orientation of the solutions suggests some analog of surface tension for localized solutions. Repelling for different fields can be compensated by this surface tension. The fact makes it possible to obtain long-lived solutions with several spatial maxima, which is proved by the numerical solutions of equations (5-7). The solutions of equations (5-7) can be interpreted as the classical variant of the model consisting of two “quarks”, or spatially separated maxima of fields.  In this case, the solutions of the equations are similar to the confinement model by flux tube [14]. Besides, model (3-4) is promising because in the Hamiltonian of the model the summand of the form $vu$ providing some confinement for fields  can be added. This is the case when, if  one or more than one field $u, v$ is different from zero, the summand  generates another field in motion equations (3-4).

%********************Bibliography.*******************************************************

\end{document}